# A trapped field of 13.4 T in a stack of HTS tapes with 30 μm substrate

A Baskys[1], K Filar[2], A Patel[1], B A Glowacki[1,3,4]

[1]Applied Superconductivity and Cryoscience Group, Department of Materials Science and Metallurgy, University of Cambridge, 27 Charles Babbage Road, Cambridge, CB3 0FS, UK

[2]International Laboratory of High Magnetic Fields and Low Temperatures, Gajowicka 95, 53-421 Wrocław, Poland

[3]Department of Physics and Energy, Bernal Institute, University of Limerick, Plassey, Ireland

[4]Institute of Power Engineering, ul. Mory 8, 01-330 Warsaw, Poland

Email: ab857@cam.ac.uk

**Abstract:**

Superconducting bulk $(RE)Ba_2Cu_3O_{7-x}$ materials (RE – rare earth elements) have been successfully used to generate magnetic flux densities in excess of 17 T. This work investigates an alternative approach by trapping flux in stacks of second generation high temperature superconducting tape from several manufacturers using field cooling and pulsed field magnetisation techniques. Flux densities of up to 13.4 T were trapped by field cooling at ~5 K between two 12 mm square stacks, an improvement of 70% over previous value achieved in an HTS tape stack. The trapped flux approaches the record values in (RE)BCO bulks and reflects the rapid developments still being made in the HTS tape performance.

## 1. Introduction

The power of electric motors and generators is linearly dependant on the magnetic flux density in the air gap between the rotor and the stator. Hence, by increasing the flux density beyond the values achievable by conventional ferromagnetic materials (1.5 T), one can increase both the power and power density of electric machine. The currently held record for the highest trapped flux density in a bulk (RE)BCO superconductor is an impressive 17.6 T by Durrell et al. [1]. Although this record value is not going to be attainable for most applications due to the geometry of the measurement (field measured between two (RE)BCO bulks in close proximity), it displays the potential and the limitations of the material. At low temperatures (20-40 K), bulk (RE)BCO is primarily limited by the mechanical properties and not critical current, which is difficult to improve, necessitating external reinforcement and application of pre-stress, as was done using a steel-band [1] or carbon fibre [2]. However, it is hoped that stacked layers of coated conductors [3–7] with strong metallic substrate will help to sustain mechanical stresses due to Lorentz force to fields above 20 T, which combined with high thermal stability makes them an attractive replacement for conventional ferromagnets in applications where high power density is critical. This work explores the current limitations of achievable trapped field using stacks of 2G HTS coated conductor tape using field cooling and pulsed field magnetisation techniques.

## 2. Experimental methods

*2.1. Samples*

Three sets of stacked tape samples were tested from Fujikura, SuperOx and SuperPower. Tape from Fujikura and Superpower were formed into stacks as loose layers compressed in the sample holder, whereas SuperOx tape was soldered into self-supporting stacks as described in [8]. The tape was cut to form 12x12 mm stacks for SuperPower and SuperOx, whereas tape from Fujikura was stacked into 10x10 mm stacks, as it was provided in a 10 mm width. Table 1 summarizes the properties of tape tested. It is evident from the data that the SuperPower HTS tape has at least twice the engineering critical current density, owing to its extremely thin substrate. Moreover, the tape from SuperPower contains artificial pinning centres, that may provide an advantage for in-field performance. To maintain a similar size of the stack, 200 layers were used instead of 100 as for the other two stacks. The 6.9 mm height of the SuperPower tape stack makes for a good comparison to previously published data on trapped field for ~65 μm thickness SuperPower tape [3]. It is worth noting that only very tiny improvements in trapped field are expected by increasing the tape stack thickness beyond 10-12 mm (the value of the tape width) [9].

Table 1. Properties of three superconducting tapes tested

| Parameter | Fujikura | SuperOx | SuperPower |
|---|---|---|---|
| Stated $I_c$ at 77 K and S.F. [A] | 450 | 400 | 450 |
| Tape width [mm] | 10 | 12 | 12 |
| Overall thickness [μm] | ~85 | ~82 (after soldering) | ~34 |
| Stabilisation | 8.2 μm Ag | 1 μm Ag, 5 μm Cu and 10 μm PbSn | 3 μm Ag |
| Engineering critical current density $J_e$ at 77 K and self-field. [kA/cm$^2$] | 52.9 | 40.7 | 107.1 |
| Number of layers per stack | 100 | 100 | 200 |
| Height of a single stack [mm] | 8.9 | 8.2 | 6.9 |

*2.2. Field cooling*

The Field cooling experiments were performed in a 15 T Oxford Instruments superconducting magnet with a 25 mm diameter bore. A custom sample holder (see Figure 1) was attached to the brass support containing a heater for temperature control. The field was measured in the central position between two tape stacks, which were separated by 2 mm, using Arepoc LHP-MP Hall sensor. Temperature was measured by a carbon ceramic temperature sensor (TMi cryogenics), placed directly on the aluminium sample holder and fixed by GE varnish.

The field cooling procedure involved heating the sample to above 100 K to ensure that the sample is no longer superconductive and applying magnetic field (larger than the expected trapped field). The sample was then cooled to the desired temperature, and the external magnetic field was ramped down to zero at a rate in the range of from 150 to 300 mT/min. The trapped flux density measurements were then taken 5 min after the end of the field cooling procedure (end of field ramp).

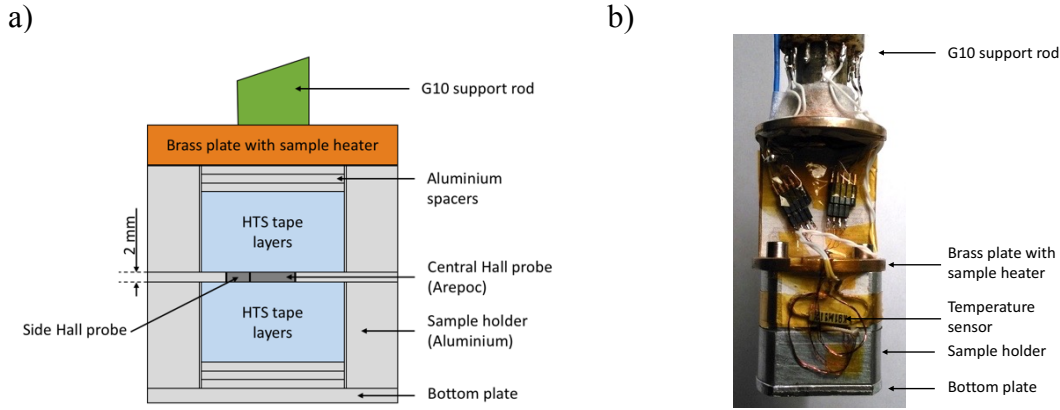

Figure 1. – Schematic diagram a) and a photograph b) of the sample holder used for field cooling experiments.

*2.3. Pulsed field magnetisation*

Pulsed field magnetisation experiments were also performed for the three different types of tape. The purpose-developed magnetisation system used is described in detail in [10]. A single stack of each tape was magnetised with the field measured 0.8 mm above the stack surface. This measurement geometry is more characteristic of the ones that will be used in applications. For pulsed field magnetisation tests, an IMRA (Iterative Magnetisation with Reducing Amplitudes) sequence with sequential cooling was used. The duration of the applied pulse was 28 ms with a rise time of 14 ms. Due to higher engineering critical current density for the SuperPower sample, much higher pulsed field magnitudes needed to be applied to fully penetrate the sample. The pulse sequence for each stack is outlined in Table 2.

Table 2. Starting flux densities $B$ in the magnetisation sequence. Pulses were applied with 0.2 T decrement until trapped flux density above the sample centre starts to decrease with additional pulses.

| Temperature [K] | Fujikura Start $B$ [T] | SuperOx Start $B$ [T] | SuperPower Start $B$ [T] |
|---|---|---|---|
| 77.4 | 2.6 | 2.6 | 2.8 |
| 60 | 4 | 4 | 6.4 |
| 50 | 4.6 | 4.6 | 6.8 |
| 40 | 5 | 5 | 7.2 |
| 30 | 5 | 5 | 7.2 |
| 20 | 4.8 | 4.8 | 7.2 |
| 10 | 4.4 | 4.4 | 7.2 |

## 3. Trapped field measurement results

*3.1. Field cooling*

The results for the field cooling of HTS tape stacks from the three manufacturers are summarised in Figure 2. The temperature in the plots below is taken to be the temperature of the sample at the end of the field ramp.

It is evident that the extremely high engineering critical current density of SuperPower 35 μm thick tape gives a significantly higher trapped flux density values. In fact, an engineering critical current density of 107.1 kA cm$^{-2}$ at 77 K and self-field is similar to values typical of bulk superconductors, even though the volume fraction of the superconductor is less than 6%. It is interesting to note that the trapped flux density is almost linear with temperature regardless of the manufacturer or the manufacturing process. The trapped field values for any temperature scale quite well with $J_e$ at 77 K and self-field (after correction by a factor of 1.2 for Fujikura tape due to smaller width). Another point to make is that the trapped field is limited by the critical current and not by mechanical properties or flux jumps down to liquid helium temperatures, unlike bulk superconductors [11]. In fact, the tapes were magnetised several times with no observable degradation.

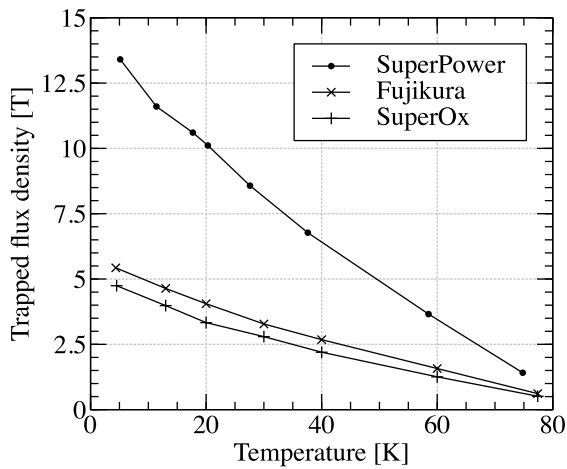

Figure 2. Field cooling results for the tree different types of HTS tape tested.

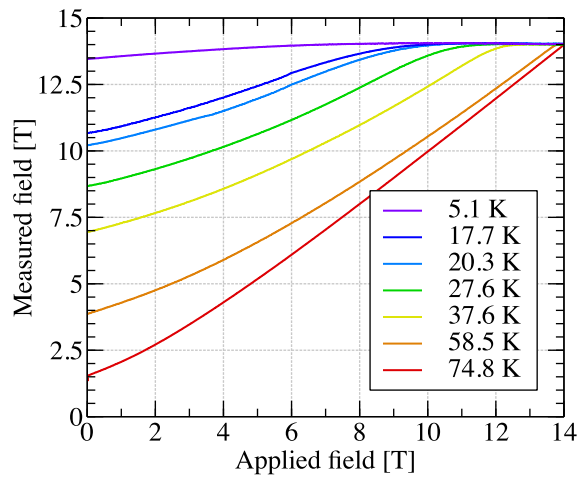

Figure 3. The measured flux density between two tape stacks (SuperPower) as the applied $B$ is ramped down from 14 T.

Figure 3 shows the field measured in between two stacks of SuperPower tape as the external magnetic field is ramped down from $\mu_0 H$ = 14 T. An applied field of 14 T is very marginal and higher magnetic field would ideally be required to be sure that the sample is fully saturated at 4.2 K.

Figure 4 shows two field cooling attempts of SuperPower tape stacks at a starting temperature of 4.2 K. Due to aggressive ramp rate used, the first attempt (Figure 4a) resulted in a flux jump with a peak temperature of more than 30 K and the flux density rapidly decreased to 5.2 T. On second attempt (Figure 4b), a slower ramp rate was used and a temperature increase was detected, however it did not result in flux avalanche and the sample managed to trap 13.4 T. It is worth noting that at the end of the ramp the sample temperature was over 5 K, and a slightly higher trapped flux density would be expected if the ramp rate was even slower and the sample was maintained at a constant 4.2 K.

The metallic substrate and highly thermally conductive stabilisation layers allowed for much faster field ramp rates than the typically used 1-2 T/h for bulk superconductors, and in the temperature range where the heat capacities are even lower. Even though a flux avalanche was observed near liquid helium temperatures, relatively fast ramp rates of 5-15 T/hr could be used for Fujikura and SuperOx samples and 10-15 T/hr for the SuperPower sample across all temperature ranges.

The trapped flux density of 13.4 T at 5.1 K or 10 T at 20 K is still lower than the record trapped field for bulk superconductors, however it is worth noting that only a fraction of identically prepared bulk (RE)BCO samples can sustain flux densities of more than 10 T due to cracks present within the sample, even when external reinforcement is applied [1,2,12]. This probability of mechanical failure for bulks at low temperatures and high fields combined with degradation due to thermal cycling [13] (when no resin impregnation is used) make the bulks less attractive for engineering applications, especially those where there is low tolerance to critical failure, regardless of the potential advantage in performance. As coated conductor manufacture matures, the availability of tape with high critical current will be increasing (even now a 450 A critical current for 12 mm width is considered average), which may eventually close the gap between maximum trapped field performance of bulk (RE)BCO and 2G HTS tape stacks. Advances in high field performance [14] and substrate reduction to 30 μm [15] or 20 μm are beneficial for many high field applications. Such advances are expected to make their way into commercially available tape in near future, resulting in even higher trapped fields. Larger width tape available from some manufacturers also provides a route to high trapped flux densities.

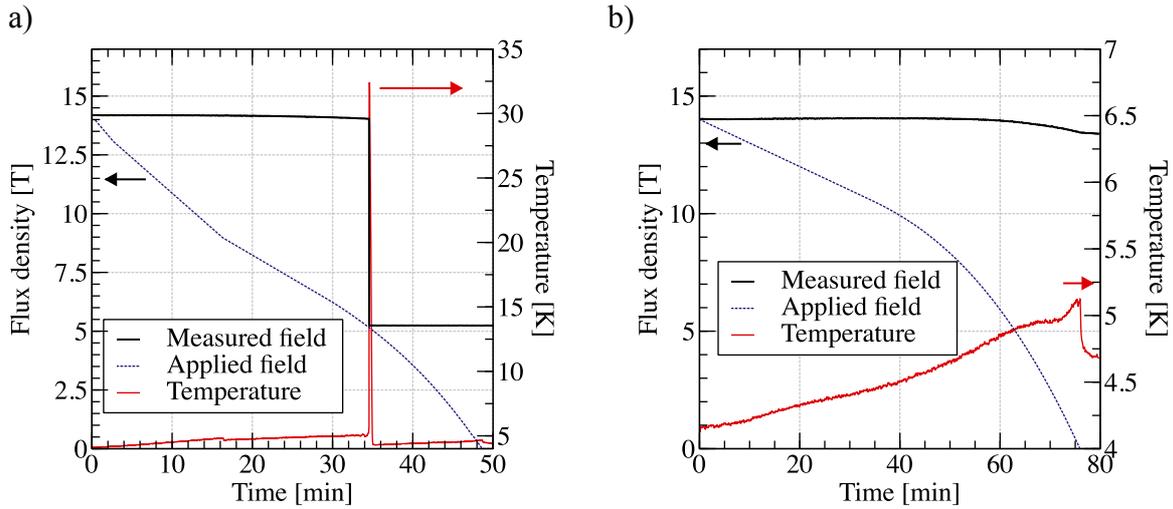

Figure 4. Field cooling of SuperPower tape stacks starting at 4.2 K. First a) and second b) ramp attempt. First ramp attempt resulted in flux jump due to fast ramp rate used.

### 3.1.1. Flux creep

Figure 5a shows the relaxation of the measured flux density between the stacks of SuperOx tape after field cooling. The measured field decay is well described by a logarithmic function after the initial ~100 s period where the decay is slower. As expected, the rate of decay decreases with temperature. Figure 5b shows the rate of decay in percent of trapped flux density per time decade for all three samples tested. However, due to limited access time to the magnet used for measurements, the creep rate for Fujikura and SuperPower samples were estimated from a smaller dataset, spanning from 3 to 5 min after the magnetisation. Moreover, at temperatures below 10 K, sample heating during the ramp exceeded 1 K, as shown in Figure 4b, but dropped down quickly after magnetisation. Therefore, the creep rate at lower temperatures may be an underestimate [16], as this in effect lowers $J/J_c$ at the operational temperature. The creep rate, however generally decreases with temperature, and the reduction is quite rapid for temperatures below 20 K. Also, interestingly, the creep rate of the SuperPower sample was the highest for high temperatures, but decreased most rapidly as the temperature was lowered. In any case, due to the logarithmic nature of the decay, it is unlikely to be an issue for motor applications for either bulk or stacked tape trapped field magnets.

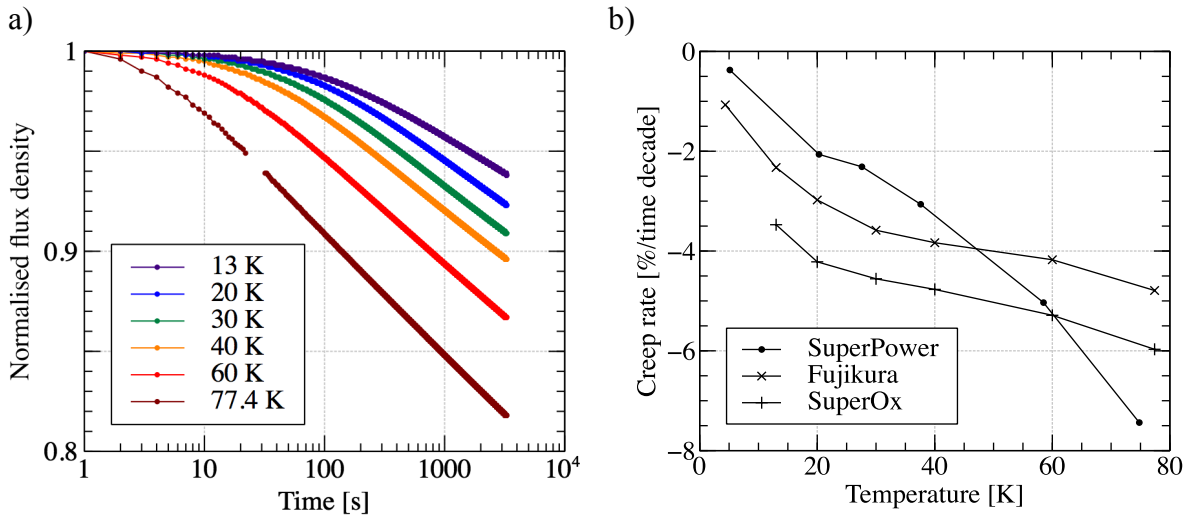

Figure 5. a) Relaxation of trapped field for SuperOx sample at different temperatures showing logarithmic decay; b) creep rates for all three types of tape tested.

*3.2. Pulsed field magnetisation*

The pulsed field magnetisation results for the three stacks are summarised in Figure 6. As expected, the SuperPower sample performed the best, followed by Fujikura, consistent with field cooling experiments. As before, at high temperatures (60-77 K) the trapped flux density ratios follow the engineering critical current density ratio at 77 K quite well, however, this is no longer true for lower temperatures. During pulsed field a magnetisation substantial amount of heat is generated, which raises the sample temperature and reduces the trapped magnetic flux.

The temperature rise is more evident at lower temperatures due to lower heat capacities at low temperatures and higher critical current. For the same reason, samples with higher $J_e$, will exhibit more heating. Therefore, increasing $J_e$ does eventually lead to diminishing results for trapped field using PFM. Giant flux jumps beneficial to magnetisation observed in high $J_e$ bulk samples [17–19] have not been observed in the tape stacks, but will be investigated using different applied pulse shapes and larger sample sizes. The flux density profiles above the SuperPower sample in Figure 7 show a large increase in trapped field as the temperature is reduced from 77 to 60 K, where the heat capacities of materials are still relatively high, but only very marginal improvements are achieved as the temperature is lowered below 40 K. Also, the profile at low temperatures does not have a conical shape, suggesting that the sample is not saturated, i.e. the current near the centre of the sample is much lower than the critical current.

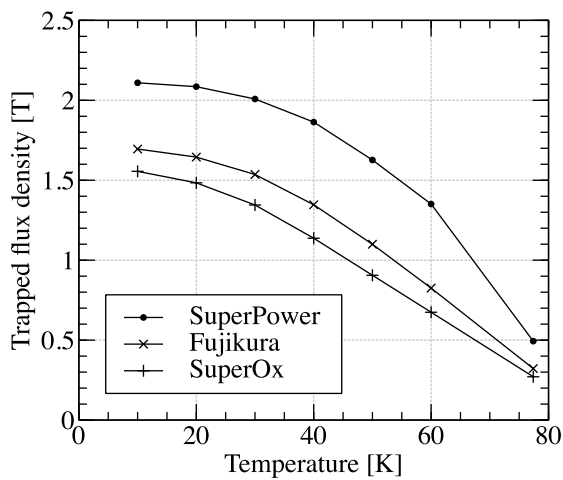 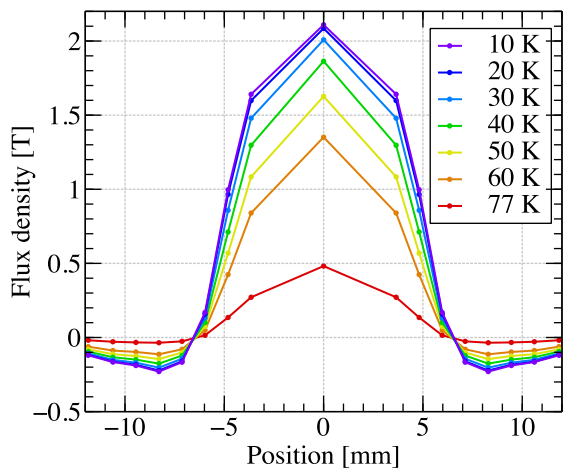

Figure 6. Pulsed field magnetisation results for a single stack of each type of tape. Field measured 0.8 mm above the sample surface. IMRA sequence with sequential cooling was used.

Figure 7. Flux density profile above the SuperPower sample after each temperature stage of the pulse sequence. The profiles at low temperatures are clearly not saturated.

## 4. Summary

Field cooling of three different types of tape was performed, showing the current state of the art for trapped field in trapped field magnets made from 2G HTS coated conductor. A maximum field of 13.4 T was achieved at ~5 K in the best sample with the highest engineering critical current density $J_e$. This is 70% higher than the previously reported [5] largest value for this type of trapped field magnet, and is approaching the record values set by bulk (RE)BCO. Also, the solder coated SuperOx tape allowed us to make solid blocks from individual layers at an expense of lower $J_e$.

No mechanical degradation was observed during multiple magnetic field ramps and temperature cycles. High thermal stability was shown by fast field ramps at very low temperatures. Unlike for bulk (RE)BCO, the performance is limited by the critical current, or rather the engineering critical current of the stack, hence the following approaches to further increase the flux density values are suggested:

- Increasing the critical current of each tape layer – larger $J_c$;
- Decreasing the overall tape thickness (most likely the substrate thickness) – larger $J_e$;
- Increasing the width of the tape, and hence the size of the stack.

On the other hand, for pulsed field magnetisation, the increase in $J_e$ does not give an equivalent increase in performance (compared to field cooling) at low temperatures, due to the increase in heating from rapid flux motion during PFM. Therefore, a reduction in temperature rise is the key point for increasing the trapped field. This can be achieved by extracting the heat more quickly by sinking the heat using thermally conductive materials, allowing more time for heat dissipation by extending the duration of the pulse itself, and optimising the applied pulse sequence. Alternatively, by tailoring the applied pulse shape flux jump assisted magnetisation can be investigated, which showed great results in PFM for bulk (RE)BCO [19].

**Acknowledgements**
The authors would like to acknowledge the financial support of EPSRC, and EraNetRus+.

**References**
[1]   Durrell J H, Dennis A R, Jaroszynski J, Ainslie M D, Palmer K G B, Shi Y-H, Campbell A M, Hull J, Strasik M, Hellstrom E E and Cardwell D A 2014 A trapped field of 17.6 T in melt-processed, bulk Gd-Ba-Cu-O reinforced with shrink-fit steel *Supercond. Sci. Technol.* **27** 82001
[2]   Tomita M and Murakami M 2003 High-temperature superconductor bulk magnets that can trap magnetic fields of over 17 tesla at 29 K. *Nature* **421** 517–20
[3]   Patel A, Filar K, Nizhankovskii V I, Hopkins S C and Glowacki B A 2013 Trapped fields greater than 7 T in a 12 mm square stack of commercial high-temperature superconducting tape *Appl. Phys. Lett.* **102** 102601
[4]   Patel A, Hopkins S C and Glowacki B A 2013 Trapped fields up to 2 T in a 12 mm square stack of commercial superconducting tape using pulsed field magnetization *Supercond. Sci. Technol.* **26** 32001
[5]   Tamegai T, Hirai T, Sun Y and Pyon S 2016 Trapping a magnetic field of 7.9 T using a bulk magnet fabricated from stack of coated conductors *Phys. C Supercond. its Appl.* **530** 20–3
[6]   Patel A, Baskys A, Hopkins S C, Kalitka V, Molodyk A and Glowacki B A 2015 Pulsed-Field Magnetization of Superconducting Tape Stacks for Motor Applications *IEEE Trans. Appl. Supercond.* **25** 5203405
[7]   Mineev N and Rudnev I 2016 Measurements and Numerical Simulations of Trapped Field in a Stack of HTS Tapes *IEEE Trans. Appl. Supercond.* **26** 8200904
[8]   Baskys A, Patel A, Hopkins S C, Kalitka V, Molodyk A and Glowacki B A 2015 Self-supporting stacks of commercial superconducting tape trapping fields up to 1.6 T using pulsed field magnetization *IEEE Trans. Appl. Supercond.* **25** 6600304
[9]   Patel A and Glowacki B A 2014 Optimisation of composite superconducting bulks made from (RE)BCO coated conductor stacks using pulsed field magnetization modelling *J. Phys. Conf. Ser.* **507** 22024
[10]  Patel A, Giunchi G, Albisetti A F, Shi Y, Hopkins S C, Palka R, Cardwell D A and Glowacki B A 2012 High Force Magnetic Levitation Using Magnetized Superconducting Bulks as a Field Source for Bearing Applications *Phys. Procedia* **36** 937–42
[11]  Krabbes G, Fuchs G, Canders W-R, May H and Palka R 2006 *High Temperature*

*Superconductor Bulk Materials* (Weinheim, FRG: Wiley-VCH Verlag GmbH & Co. KGaA)

[12]   Ren Y, Weinstein R, Liu J, Sawh R P and Foster C 1995 Damage caused by magnetic pressure at high trapped field in quasi-permanent magnets composed of melt-textured YBaCuO superconductor *Phys. C Supercond. its Appl.* **251** 15–26

[13]   Tomita M and Murakami M 2002 Mechanical properties of bulk superconductors with resin impregnation *Supercond. Sci. Technol.* **15** 808–12

[14]   Selvamanickam V, Gharahcheshmeh M H, Xu A, Zhang Y and Galstyan E 2015 Critical current density above 15 MA cm$^{-2}$ at 30 K, 3 T in 2.2 μm thick heavily-doped (Gd,Y)Ba$_2$Cu$_3$O$_x$ superconductor tapes *Supercond. Sci. Technol.* **28** 72002

[15]   Sundaram A, Zhang Y, Knoll A R, Abraimov D, Brownsey P, Kasahara M, Carota G M, Nakasaki R, Cameron J B, Schwab G, Hope L V, Schmidt R M, Kuraseko H, Fukushima T and Hazelton D W 2016 2G HTS wires made on 30 μm thick Hastelloy substrate *Supercond. Sci. Technol.* **29** 104007

[16]   Weinstein R, Liu J, Ren Y, Sawh R, Parks D, Foster C and Obot V 1996 No Title *Proceedings 10th Anniversary HTS Workshop on Physics, Materials and Applications,* ed B Batlogg, C W Chu, W K Chu, D U Gubser and K A Muller (World Scientific, Singapore) p 625

[17]   Fujishiro H, Tateiwa T, Fujiwara A, Oka T and Hayashi H 2006 Higher trapped field over 5T on HTSC bulk by modified pulse field magnetizing *Phys. C Supercond.* **445–448** 334–8

[18]   Weinstein R, Parks D, Sawh R, Carpenter K and Davey K 2016 Anomalous results observed in magnetization of bulk high temperature superconductors - a windfall for applications *J. Appl. Phys.* **119** 133906

[19]   Zhou D, Ainslie M D, Shi Y, Dennis A R, Huang K, Hull J R, Cardwell D A and Durrell J H 2017 A portable magnetic field of >3 T generated by the flux jump assisted, pulsed field magnetization of bulk superconductors *Appl. Phys. Lett.* **110** 62601